\documentclass[aps,prl,10pt,a4paper,twocolumn,footinbib,superscriptaddress,showkeys]{revtex4-2}
\usepackage[utf8]{inputenc}
\usepackage{fancyhdr}
\usepackage{amsmath}
\usepackage{amssymb}
\usepackage{amsthm}
\usepackage{graphicx}
\usepackage{dsfont}
\usepackage{makecell}
\usepackage{yfonts}
\usepackage{MnSymbol}
\usepackage{physics}
\usepackage{xcolor}
\usepackage{balance}
\usepackage[colorlinks=true,citecolor=blue,linkcolor=blue,urlcolor=blue]{hyperref}

\setcellgapes{3pt}
\makegapedcells

\begin{document}
\bibliographystyle{apsrev4-2}
\setcounter{secnumdepth}{2}

\title{Probabilistic generation of two-mode binomial cat states using cross-Kerr interactions}

\author{S. Zhao}
\affiliation{Institute for Quantum Materials and Technology, Karlsruhe Institute of Technology, 76344 Eggenstein-Leopoldshafen, Germany}

\author{A. Metelmann}
\affiliation{Institute for Quantum Materials and Technology, Karlsruhe Institute of Technology, 76344 Eggenstein-Leopoldshafen, Germany}
\affiliation{Institute for Theory of Condensed Matter, Karlsruhe Institute of Technology, 76131 Karlsruhe, Germany}
\affiliation{Institut de Science et d’Ingénierie Supramoléculaires (ISIS, UMR7006), University of Strasbourg and CNRS}

\author{S. Qvarfort}
\affiliation{Nordita, KTH Royal Institute of Technology and Stockholm University, Hannes Alfv\'{e}ns v\"{a}g 12, SE-106 91 Stockholm, Sweden}
\affiliation{Department of Physics, Stockholm University, AlbaNova University Center, SE-106 91 Stockholm, Sweden}

\begin{abstract}
Superpositions of macroscopically distinct coherent states, or cat states, are a key resource for quantum technologies. In particular, two-mode binomial cat states can give rise to exact quantum error correction in continuous-variable quantum computing. However, their preparation typically requires non-Gaussian initial states, such as Fock or NOON states, which are challenging to realize experimentally. Here, we propose a protocol to generate two-mode binomial cat states where the only requirements are Gaussian initial states in combination with heralded heterodyne measurements. Our approach utilizes cross-Kerr interactions between bosonic modes commonly realizable in superconducting circuit architectures. We show that by adjusting the input state parameters, our protocol remains robust to dissipation under realistic experimental conditions. Our work provides a practical route to realizing multinomial cat states in current experimental platforms without requiring non-Gaussian initial resources, overcoming a key limitation of existing preparation schemes.

\end{abstract}

\maketitle

\section{Introduction}\label{sec_introduction}
The ability to realize quantum superpositions of macroscopically distinct states, or cat states, is central to quantum technologies such as computing~\cite{BartlettPRA02, ChaeNC24, DyakonovFTiM13, JeongPRA02, Ralph01, SoodACME24}, metrology~\cite{MunroPRA02, ShimizuPRL05, TatsutaPRA19}, and communication~\cite{vanEnkPRA01, JeongPRA01}. While single-mode cat states (first proposed in 1986~\cite{YurkePRL86}) have been widely studied, multimode realizations have attracted increasing attention as they exhibit richer structures and enhanced robustness~\cite{BergmannPRA16, MalekiEPJP21, GroiseauPRA21, AlbertQST19, GertlerPQ23}. A promising candidate is the two-mode bosonic representation of spin cat states~\cite{MichaelPRX16}, known as a two-mode binomial cat state, which exhibits binomial excitation-number statistics. A key feature of this state is that its constituent coherent states are fully orthogonal, which in turn enables exact quantum error correction~\cite{AlbertPRA18}. This is in contrast to single-mode bosonic cat states, where overlaps vanish only asymptotically, and the error correction becomes necessarily approximate.

The experimental realization of two-mode binomial cat states remains an active area of research. The standard approach is based on the Kerr evolution of an initial two-mode binomial coherent state into its corresponding binomial cat states~\cite{YurkePRL86, KeatingPRL16, BarzanjehPRA16}. While the initial two-mode binomial coherent states can be directly prepared in atomic and spin ensembles using a laser drive~\cite{AgarwalPRA97, MalekiEPJP21, MalekiJOSABJ20a, LoFrancoPLA10a}, their realization in bosonic systems can be difficult. This is because, in atomic or spin platforms, the finite-dimensional Hilbert space naturally fixes the maximum excitation number, whereas in bosonic systems, it corresponds to the total excitation number across the bosonic modes, which is unbounded and cannot be easily fixed. To integrate with the versatile toolbox of bosonic quantum technologies, existing proposals for two-mode bosonic systems therefore typically require access to nontrivial and highly non-Gaussian initial states with a fixed total excitation number, such as a specific Fock or NOON state~\cite{BarzanjehPRA16, AlbertPRA18, MalekiEPJP21}. An alternative and complementary approach is provided by dissipative protocols, where carefully tailored dissipation can continuously steer the system from an initial vacuum state toward the desired two-mode binomial cat state~\cite{Zhao26}.

In this work, we propose a probabilistic protocol for the preparation of two-mode binomial cat states in bosonic systems. The protocol makes use of cross-Kerr interactions as well as only Gaussian input states and measurements, in contrast to existing approaches that rely on non-Gaussian initial states. We utilize the pairwise cross-Kerr interactions in a three-mode bosonic system. With each mode being initialized in a single-mode coherent state, which is cheap to prepare, a subsequent projective measurement on the auxiliary (third) mode then heralds the formation of a two-mode binomial cat state in the other two modes. Moreover, we show that in an open environment with single-photon loss, there exists an optimal choice of the initial coherent amplitude for the acilla mode, which maximizes the protocol fidelity. Finally, by considering state-of-the-art parameters for superconducting devices ~\cite{BlaisRMP21, ZoepflPRL23}, we find that our protocol can be readily implemented. More broadly, our work demonstrates how readily available Gaussian states and naturally occurring Kerr-type nonlinearities can be combined to generate multinomial cat states, providing an experimentally accessible route towards their realization in bosonic quantum platforms.

\section{Cross-Kerr Hamiltonian}\label{sec_cross_kerr_hamiltonian}
The central ingredient of the proposed protocol is the cross-Kerr nonlinear interaction, which induces an energy shift proportional to the product of the excitation numbers of two modes. This effect has been utilized to perform various quantum tasks such as entanglement preparation~\cite{HePRA09, JeongPRA05, LiaoJPBAMOP06a, JosseJOBQSO04}, photon counting~\cite{DingPRL17}, and the design of nonlinear quantum devices~\cite{BittencourtPRB23, QianPRA21}. Owing to its versatility, the cross-Kerr interaction has been implemented in a variety of physical platforms, including superconducting circuits~\cite{BourassaPRA12a}, optomechanical systems~\cite{SaikoJL21}, Rydberg gases~\cite{SinclairPRR19}, four-level atomic systems~\cite{SinclairPRA07}, and magneto-optical systems~\cite{RahmanRiP23}.

In superconducting circuit implementations utilizing nonlinear elements 
based on Josephson junctions~\cite{JosephsonPL62, FrattiniAPL17a, LescanneNP20}, both self- and cross-Kerr interactions are typically generated simultaneously (see appendix~\ref{sec_app_self_cross}). Setting $\hbar=1$, the system Hamiltonian is given by \begin{align}
    \hat{H} &=\hat{H}_{\mathrm{sk}}+\hat{H}_{\mathrm{ck}}, \label{Eq02}
\end{align}
where \begin{align}
    \nonumber \hat{H}_{\mathrm{sk}}&=g_a(\hat{a}^\dag\hat{a})^2+g_b(\hat{b}^\dag\hat{b})^2+g_c(\hat{c}^\dag\hat{c})^2,\\
    \hat{H}_{\mathrm{ck}}&=g_{ab}\hat{a}^\dag\hat{a}\hat{b}^\dag\hat{b}+g_{ac}\hat{a}^\dag\hat{a}\hat{c}^\dag\hat{c}+g_{bc}\hat{b}^\dag\hat{b}\hat{c}^\dag\hat{c}, \label{eq_sk_ck}
\end{align}
which describe the self-Kerr and cross-Kerr nonlinearities respectively, and where $\hat{a}$, $\hat{b}$, and $\hat{c}$ are bosonic annihilation operators associated with the three resonator modes, and $g_\bullet$ are self- and cross-Kerr coefficients. Here, we included self-Kerr nonlinearities as they naturally accompany cross-Kerr interactions in superconducting circuits, and their presence can enhance the protocol.

\section{Protocol}\label{sec_protocol}
We now show how the cross-Kerr nonlinearities can be used to generate two-mode binomial cat states. The constituent two-mode binomial coherent states of such cat states are denoted as~\cite{PerelomovSPU77} \begin{align}
    |N,\xi\rangle&=\frac{1}{\sqrt{\mathcal{N}}}\sum_{n=0}^N\sqrt{\binom{N}{n}}\xi^n|n\rangle\otimes|N-n\rangle, \label{eq_def_su2_state}
\end{align}
where $\mathcal{N}=(1+|\xi|^2)^N$ is the normalization factor, $\binom{N}{n}=N!/(n!(N-n)!)$ is the binomial coefficient, $\xi\in\mathds{C}$ is the coherent amplitude, $|n\rangle\otimes|N-n\rangle$ is the tensor product of Fock states $|n\rangle$ and $|N-n\rangle$, and $N\in\mathds{N}$ is the maximum excitation number in the system. With this, the target two-mode binomial cat state is defined as \begin{align}
    |\psi(N,\xi)\rangle&=\frac{1-i}{2}|N,\xi\rangle+\frac{1+i}{2}|N,-\xi\rangle. \label{eq_su2_cat_state}
\end{align}
Ideally, we require $N$ to be sufficiently large, so that the cat components $|N,\pm\xi\rangle$ remain macroscopic, as well as $|\xi|=1$ to ensure the cat components remain orthogonal~\cite{Zhao26}. Both conditions together fulfill the definition of a cat states as a superposition of two macroscopically distinct states.

To construct the protocol, we exploit the expansion of a product of initial single-mode coherent states in the two-mode binomial coherent state basis~\cite{klimov2009group} (see appendix~\ref{sec_app_detailed_calculation}), \begin{align}
    |\alpha\rangle\otimes|\beta\rangle=\sum_{k=0}^\infty \sqrt{P(k)} |k,\xi\rangle, \label{eq_conversion}
\end{align}
where $|\alpha\rangle$ and $|\beta\rangle$ are single-mode bosonic coherent states with their amplitudes $\alpha,\beta$ assumed to be real without loss of generality, $\xi=\frac{\alpha}{\beta}$, and $P(k)=e^{-\lambda}\lambda^k/k!$ is the Poisson distribution with the mean $\lambda=|\alpha|^2+|\beta|^2$. This decomposition demonstrates that the product of two single-mode coherent states can be represented as an infinite superposition of two-mode binomial coherent states. Consequently, a projection onto $k=N$ prepares the initial two-mode binomial coherent state required for the Kerr-cat protocol~\cite{YurkePRL86, XuPRX25}. The generation of the target state $|\psi(k,N)\rangle$ then requires (i) the self-Kerr nonlinear evolution for each fixed $k$ into a cat state, and (ii) selection of a particular $k$ that corresponds to the total excitation number in the first two modes. Both requirements are implemented by coupling to an auxiliary mode initialized in another single-mode coherent state $|\gamma\rangle$ via the Hamiltonian Eq.~\eqref{Eq02} with $g_{ac}=g_{bc}$ (achievable via flux tuning of the circuit elements). That is, the initial state $|\alpha\rangle\otimes|\beta\rangle\otimes|\gamma\rangle$ evolves as (see appendix~\ref{sec_app_detailed_calculation})\begin{align}
    |\Psi(t)\rangle&=\sum_{k=0}^\infty c_k e^{-i\mathcal{K}_2(\hat{a}^\dag\hat{a})^2t}|k,\xi_k\rangle \otimes e^{-ig_c(\hat{c}^\dag\hat{c})^2t}|\gamma_k\rangle, \label{Eq12}
\end{align}
where $c_k=\sqrt{P(k)} e^{-ig_bk^2t}$, $\xi_k=e^{-i(g_{ab}-2g_b)kt}\alpha/\beta$, $\gamma_k=e^{-ig_{ac}k t}\gamma$, and $\mathcal{K}_2=g_a+g_b-g_{ab}$ is the effective self-Kerr strength. Since $e^{-ig_b k^2 t}$ is merely a relative phase between the components indexed by $k$, it does not affect the protocol that ultimately aims to project the state onto a particular $k$, and thus will be ignored. At the target time $t_0=\pi/(2|\mathcal{K}_2|)$~\cite{YurkePRL86}, the system evolves into a superposition of the target binomial cat states $|\psi(k,\xi)\rangle$ (see appendix~\ref{sec_app_detailed_calculation}) \begin{align}
    |\Psi(t_0)\rangle&=\sum_{k=0}^\infty \sqrt{P(k)} |\psi(k,\xi_k)\rangle\otimes e^{-ig_c(\hat{c}^\dag\hat{c})^2t_0}|\gamma_k\rangle. \label{Eq03}
\end{align}
Crucially, the total excitation number $k$ of the system modes ($a$ and $b$) is now mapped onto a phase in the auxiliary mode. 

\begin{figure}
    \centering
    \includegraphics[width=0.85\linewidth]{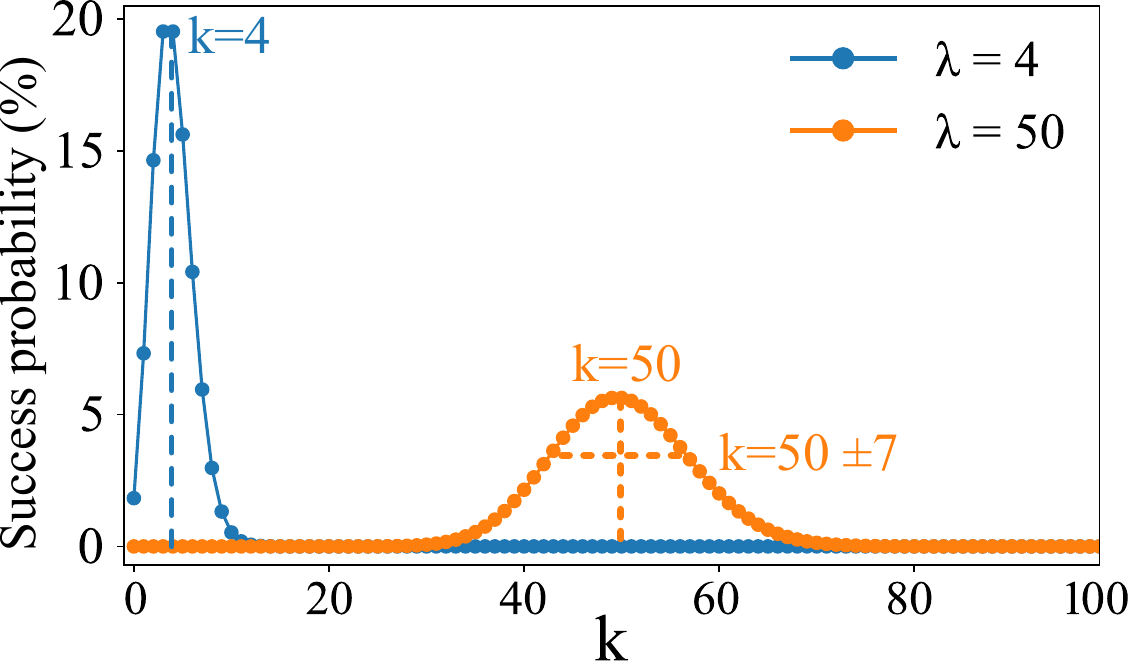}
    \caption{Success probability $P(k)$ (from Eq.~\eqref{eq_conversion}) of the protocol in the ideal case, for $\lambda=4$ and $\lambda=50$. The vertical dashed lines indicate the most probable value of $k$, while the horizontal dashed line marks the range of $k$ considered as successful outcomes. Although other sufficiently large values of $k$ outside this range also constitute successful outcomes in principle, their probabilities carry much smaller weights, which in turn results in a lower-fidelity on the heralded cat state.}
    \label{fig:poisson_comparison}
\end{figure}

Now, by collapsing the superposition state in Eq.~\eqref{Eq03} onto a single index $k$, we effectively prepare a two-mode binomial cat state in the system modes. To do so, we perform a heterodyne measurement on the auxiliary mode, which essentially projects it onto a coherent state $\hat M_\phi = |\phi\rangle\langle \phi|/\pi$~\cite{WisemanQSO96}. Each measurement yields a complex outcome $\phi\in\mathds{C}$, and we can regard the measurement as successful in realizing the desired projector $\hat{M}_k=|\gamma_k\rangle\langle\gamma_k|$, whenever $\phi$ lies within a sufficiently small neighborhood $S=\{\phi \mid |\phi-\gamma_k|\ll\delta\}$ of the target value $\gamma_k$. With this, a successful heterodyne measurement effectively selects a single value of $k$ from the superposition in Eq.~\eqref{Eq03}, thereby heralding the corresponding cat state $|\psi(k,\xi_k)\rangle$ in the system modes. The corresponding success probability is
\begin{align}
    \nonumber \mathrm{Prob}(k)&=\sum_{k'=0}^\infty P(k') \int_{S} \frac{d^2\phi}{\pi}|\langle \phi|e^{-ig_c(\hat{c}^\dag\hat{c})^2t_0}|\gamma_{k'}\rangle|^2,\\
    &\approx \delta^2\sum_{k'=0}^\infty P(k') |\langle \gamma_k|e^{-ig_c(\hat{c}^\dag\hat{c})^2t_0}|\gamma_{k'}\rangle|^2\label{eq_probability_full}
\end{align}
where in the second line, we approximated the integral as an multiplication by the area of $S$ for sufficiently small $\delta$. For simplicity, in the remainder of this work, we rescale the probability by the factor $\delta^2$ and instead work with the corresponding probability density, as is natural for continuous-variable systems: \begin{align}
    Q_\rho(k)&=\frac{\mathrm{Prob}(k)}{\delta^2}=\sum_{k'=0}^\infty P(k')|\langle \gamma_k|e^{-ig_c(\hat{c}^\dag\hat{c})^2t_0}|\gamma_{k'}\rangle|^2. \label{eq_probability_ideal_case}
\end{align}
This quantity is the renormalized Husimi-$Q$ function, $Q_\rho(k)=\pi Q(\gamma_k)$, obtained by evaluating the Husimi-$Q$ function $Q(\phi)$ at $\phi=\gamma_k$. For convenience, we will continue to refer to $Q_\rho(k)$ as the success probability throughout the remainder of this work, since it is essentially the actual success probability rescaled by the area of an artificially chosen region $S$. The post-measurement state of the two target modes corresponding to the ideal outcome $\phi=\gamma_k$ is then given by
\begin{align}
    |\Psi(t_0)\rangle_\phi&=\frac{1}{\sqrt{Q_\rho(k)}}\sum_{k'\in D}\sqrt{P(k')}|\psi(k',\xi_{k'})\rangle\otimes |\gamma_{k'}\rangle.\label{eq_post_state_full}
\end{align}
Eq.~\eqref{eq_post_state_full} highlights that the heralded state is, in general, not associated with a unique target state $|\psi(k,\xi_k)\rangle$. This originates from the $2\pi$ periodicity of the auxiliary coherent-state phase, which implies that different excitation numbers can correspond to the same coherent state, i.e., $|\gamma_k\rangle=|\gamma_{k'}\rangle$ for $k'\in D=\{k'\mid k-k'=2\pi/(g_{ac}t_0)\}$. As a result, the heralded state may still contain contributions from multiple values of $k'$. Nevertheless, these additional contributions can be made negligible: as illustrated in Fig.~\ref{fig:poisson_comparison}, by post-selecting a target value $k$ such that all other values $k'\in D$ lie in the tails of the Poisson distribution $P(k)$, the corresponding terms acquire only very small statistical weights $P(k')$. Their contributions to the heralded state are therefore strongly suppressed and can be safely neglected. Consequently, the superposition state in Eq.~\eqref{eq_post_state_full} is well approximated by the target state associated with a single excitation index $k$. This allows Eq.~\eqref{eq_post_state_full} to be approximated by \begin{align}
    |\Psi(t_0)\rangle\approx \sqrt{\frac{P(k)}{Q_\rho(k)}}|\psi(k,\xi_{k})\rangle\otimes |\gamma_{k}\rangle, \label{eq_post_state_full_02}
\end{align}
which is not yet properly normalized due to the prefactor $\sqrt{P(k)/Q_\rho(k)}$.

Moreover, in the ideal regime, where the success probability $Q_\rho(k)$ is maximized while maintaining a clear projection onto a single value of $k$, two conditions must be satisfied. First, the fine-tuning of $g_c=2m\pi/t_0$ for some integer $m$, so that the self-Kerr evolution at $t_0$ effectively behaves as the identity operator, i.e., $e^{-ig_c(\hat{c}^\dag\hat{c})^2t_0}|n\rangle=e^{-2imn\pi}|n\rangle=|n\rangle$. Second, a sufficiently large amplitude $\gamma$, such that other summands with $k'\neq k$ vanishes, i.e., $|\langle \gamma_k |\gamma_{k'}\rangle|^2=\delta_{k,k'}$. With these conditions, the dominant contribution to $Q_\rho(k)$ comes from the target value $k$, which allows the approximation $Q_\rho(k)\approx P(k)$, and thus Eq.~\eqref{eq_post_state_full_02} becomes properly normalized and reduces to \begin{align}
    |\Psi(t_0)\rangle&\approx |\psi(k,\xi_{k})\rangle\otimes |\gamma_{k}\rangle.
\end{align}
In particular, for larger $\lambda=|\alpha|^2+|\beta|^2$, although the probability $P(k)$ of obtaining any specific value of $k$ decreases, the Poisson distribution broadens, making a wider range of $k$ values available as successful outcomes. Importantly, our protocol does not require a specific value of $k$; rather, any sufficiently large $k$ would correspond to a suitable cat state, and can be considered as a successful outcome. Hence, the actual success probability becomes the cumulative probability obtained by summing $P(k)$ over all such suitable values of $k$. This total probability remains significant even for large $\lambda$. For example, within one standard deviation of the mean, $\lambda-\sqrt{\lambda}\leq k\leq\lambda+\sqrt{\lambda}$, the cumulative probability approaches $68\%$ as $\lambda$ becomes increasingly large (see Fig.~\ref{fig:poisson_comparison}).

As a brief summary, the protocol consists of: \begin{enumerate}
    \item initializing the system in $|\alpha\rangle\otimes|\beta\rangle\otimes|\gamma\rangle$ where $\gamma\gg 1$ is ideally as large as possible,

    \item letting the system evolve under the Hamiltonian in Eq.~\eqref{eq_sk_ck} for $g_{ac}=g_{bc}$ and $g_c=2m\pi/t_0$, and

    \item performing a heterodyne measurement on the auxiliary mode at time $t_0=\pi/(2|\mathcal{K}_2|)$.
\end{enumerate}
The heralding is considered successful provided that $k$ is sufficiently large to yield a suitable cat state with near-unit fidelity. The details for realizing the required parameter relations $g_{ac}=g_{bc}$ and $g_c=2\pi/t_0$ ($m=1$) in superconducting systems can be found in appendix~\ref{sec_app_self_cross}.

\section{Impact of dissipation}\label{sec_impact_of_dissipation}
All quantum systems inevitably couple to their environment, and as a result, the ideal protocol can only be approximately realized. In superconducting circuits, the dominant noise channel is single-photon loss, which gives rise to both energy relaxation and effective dephasing. A complete open-system analysis including thermalization and dephasing is presented in appendix~\ref{sec_app_open_system_evolution}, following a similar approach to Ref.~\cite{MogilevtsevPRA09}. 

In the following, we derive a condition for mitigating the impact of single-photon loss, which effectively imposes an upper bound on the optimal choice of $\gamma$. The Lindblad master equation incorporating single-photon loss is
\begin{align}
    \dot{\hat{\rho}}=-i[\hat{H},\hat{\rho}]+\sum_{\eta=a,b,c}\hat{\mathcal{D}}(\sqrt{\kappa_{\eta}}\hat{\eta})\hat{\rho},
    \label{eq_lindblad}
\end{align}
where $\hat{H}$ is given in Eq.~\eqref{eq_sk_ck}, $\kappa_{\eta}$ is the single-photon loss rate in the mode $\eta\in\{a,b,c\}$, and $\hat{\mathcal{D}}(\hat{o})\hat{\rho}=\hat{o}\hat{\rho}\hat{o}^\dag-\frac{1}{2}(\hat{o}^\dag\hat{o}\hat{\rho}+\hat{\rho}\hat{o}^\dag\hat{o})$ is the usual Lindblad dissipator. At the target time $t_0$, the initially pure state $|\alpha\rangle\otimes|\beta\rangle\otimes|\gamma\rangle$ evolves into
\begin{align}
    \nonumber \hat{\rho}(t_0)&=\sum_{k_1,k_2}  f(\hat{F}_-)\sqrt{P'(k_1)P'(k_2)}|\psi(k_1,\xi_{k_1}')\rangle\langle\psi(k_2,\xi_{k_2}')|\\
    &\qquad \otimes|\gamma_{k_1}'\rangle\langle \gamma_{k_2}'|, \label{Eq07}
\end{align}
where by defining the damped coherent amplitude $\nu_\eta'=\nu_\eta e^{-\kappa_\eta t_0/2}$ for $\nu_{\eta\in\{a,b,c\}}=\alpha,\beta,\gamma$, the primed symbols are similarly defined. The operator-valued factor $f(\hat{F}_-)$ is defined as 
\begin{align}
    f(\hat{F}_-)=\prod_{\eta=a,b,c}e^{(\hat{F}_{-}^{(\eta)}-1+e^{-\kappa_{\eta} t})|\nu_\eta|^2},\label{EqOVfactor}
\end{align}
where $\hat{F}_{-}^{(\eta)}=\frac{\kappa_\eta}{\hat{A}_\eta+\kappa_{\eta}}(1-e^{-(\hat{A}_{\eta}+\kappa_{\eta})t})$ accounts for dissipation, $\hat{A}_{\eta}=2ig_\eta\hat{K}_{0}^{(\eta)}+i\sum_{\varsigma\neq\eta}g_{\eta\varsigma}\hat{K}_{0}^{(\varsigma)}$ describes the unitary evolution, and $\hat{K}_0^{(\eta)}(\hat{\rho})=\hat{\eta}^\dag\hat{\eta}\hat{\rho}-\hat{\rho}\hat{\eta}^\dag\hat{\eta}$ are superoperators that capture the underlying symmetry of the dynamics~\cite{AlbertPRA14a}. The factor in Eq.~\eqref{EqOVfactor} captures the additional decoherence induced by single-photon loss, which reduces both the success probability Eq.~\eqref{eq_probability_special_case} and fidelity Eq.~\eqref{eq_fidelity_special_case} of the protocol. Compared to the ideal evolution in Eq.~\eqref{Eq03}, we observe that single-photon loss induces both a change in the statistical weight $P(k)\to P'(k)$ via amplitude damping, and decoherence via the $f(\hat{F}_-)$ factor. As a result, the success probability for heralding the cat state $|\psi(k,\xi_k)\rangle$ becomes
\begin{align}
    \nonumber Q_\rho(k)&=\mathrm{Tr}[\hat{M}_k\hat{\rho}(t_0)\hat{M}_k^\dag]\\
    &=\sum_{\substack{n_1,n_3\\ m_3,k_1}}f_\mathrm{prob}(n_3,m_3)g_\mathrm{prob}(k,k_1,n_1,n_3,m_3), \label{eq_probability_special_case}
\end{align}
where $f_\mathrm{prob}$ is $f(\hat{F}_-)$ evaluated by substituting the superoperators $\hat{K}_0^{(\bullet)}$ with scalars, i.e., $\hat{K}_{0}^{(a)}=\hat{K}_{0}^{(b)}=0$ and $\hat{K}_{0}^{(c)}=n_3-m_3$, and $g_\mathrm{prob}$ is a product of several Poisson and binomial probability amplitudes defined in appendix~\ref{subsec_probability_and_fidelity}.

The fidelity of the heralded two-mode binomial cat state $|\psi(k,\xi)\rangle$ in the presence of noise is therefore
\begin{align}
    \nonumber \mathcal{F}(k)&=\frac{\langle\psi(k,\xi_k)|\mathrm{Tr}_c[\hat{M}_k\hat{\rho}(t_0)\hat{M}_k^\dag]|\psi(k,\xi_k)\rangle}{Q_\rho(k)}\\
    \nonumber &=\frac{P'(k)}{Q_\rho(k)}\sum_{\substack{n_1,n_3\\ m_1,m_3}}f_\mathrm{fid}(n_1,m_1,n_3,m_3)\\
    &\hspace{3cm} \times g_\mathrm{fid}(k,n_1,n_3,m_1,m_3), \label{eq_fidelity_special_case}
\end{align}
where $f_\mathrm{fid}$ is obtained from $f(\hat{F}_-)$ by substituting $\hat{K}_{0}^{(a)}=n_1-m_1=-\hat{K}_0^{(b)}$ and $\hat{K}_{0}^{(c)}=n_3-m_3$, and similarly, $g_\mathrm{fid}$ is another product of several probability amplitudes defined in appendix~\ref{subsec_probability_and_fidelity}. In the absence of dissipation, $f_\mathrm{fid}\to 1$, the summations factorize into independent terms with each being summed to unity, and Eq.~\eqref{eq_fidelity_special_case} reduces to $\mathcal{F}(k)=P(k)/Q_\rho(k)\approx 1$ as expected.

The summand in both Eq.~\eqref{eq_probability_special_case} and Eq.~\eqref{eq_fidelity_special_case} consist of two contributions: the decoherence factors $f_\bullet$ and the distribution products $g_\bullet$. While the former imposes an upper bound on the optimal choice of $\gamma$ as discussed below, the latter accounts for the mismatch between the corresponding distribution amplitudes, which has an negligible effect compared with that of the former (see appendix~\ref{sec_app_further_modifications}).

\begin{figure}
    \centering
    \includegraphics[width=\linewidth]{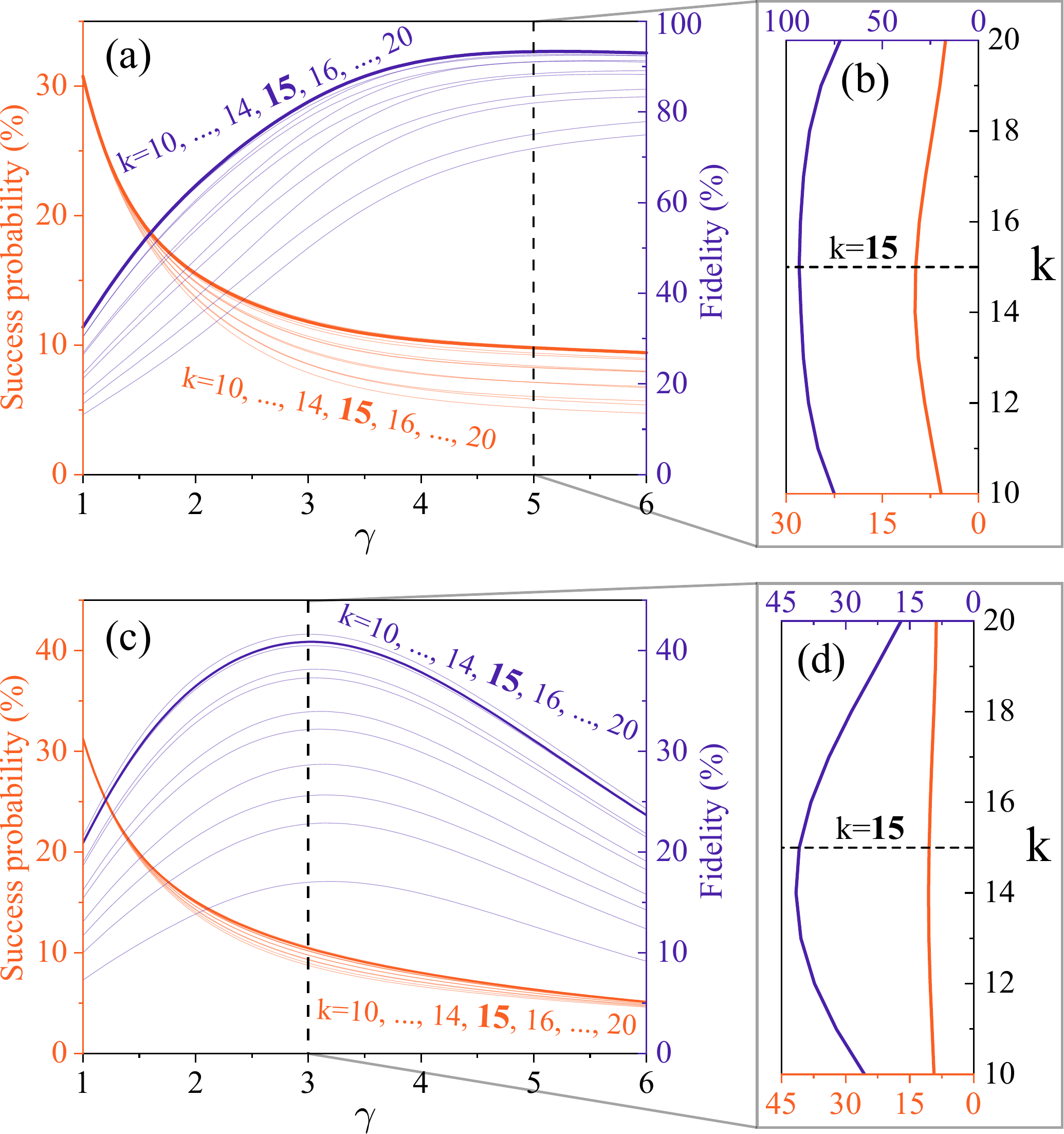}
    \caption{Fidelity and success probability of the protocol in the presence of single-photon loss, using $g_{ac}=g_{bc}=\sqrt{8}g_{ab}$, $g_a=g_b=g_c/8$, $g_{ac}=4\sqrt{8}g_a$ and $t_0=\pi/(4g_a)$ (see appendix~\ref{sec_app_self_cross}), and initial amplitudes $\alpha=\beta=\sqrt{7.5}$. (a) and (c): Fidelity and probability versus $\gamma$ with the decay rates (a) $\kappa_a=0.001$, $\kappa_b=0.003$, $\kappa_c=0.005$, and (c) $\kappa_a=0.03$, $\kappa_b=0.05$, $\kappa_c=0.07$. (b) and (d): Fidelity and probability versus $k$ evaluated at the optimal $\gamma$ (vertical dashed lines in (a),(c)). Due to computational limitations, the larger decay rates used in (c) were chosen to illustrate the qualitative behavior of (a) at larger values of $\gamma \geq 6$, including the emergence of a concave curve.}
    \label{fig:prob_fid_plot}
\end{figure}
To mitigate the effect of dissipation, we optimize the initial coherent amplitude $\gamma$ of the auxiliary mode. In the ideal (noise-free) case, $\gamma$ is required to be as large as possible to enforce the orthogonality condition $|\langle\gamma_k|\gamma_{k'}\rangle|^2=\delta_{k,k'}$. However, in the presence of single-photon loss ($\kappa_c\neq 0$), increasing $\gamma$ also enhances the decoherence mediated via $f_\mathrm{fid}$ in Eq.~\eqref{eq_fidelity_special_case}, which leads to an overall reduction in fidelity. As a result, there exists an optimal value of $\gamma$ that balances these competing effects: larger $\gamma$ improves the selectivity of the measurement, while smaller $\gamma$ catches less severe decoherence in the system. 

To demonstrate the trade-off in initial-state amplitude, we consider experimentally realistic parameters for superconducting devices. We choose decay rates $\kappa_{a}=0.001g_a$, $\kappa_{b}=0.003g_a$, and $\kappa_{c}=0.005g_a$~\cite{BlaisRMP21, ZoepflPRL23}, where self- and cross-Kerr interactions are typically on the order of MHz, and single-photon loss rates are on the order of kHz. Fig.~\ref{fig:prob_fid_plot}(a) shows the existence of an optimal $\gamma\approx 5$, around which we achieve fidelity of $\sim95\%$ near $k=15$ and above $\sim75\%$ within $k\pm 5$. For a target fidelity of $90\%$ across $13\leq k\leq 17$, the total success probability is $46.45\%$. To make the existence of such an optimal $\gamma$ more pronounced, we simulate the same dynamics with significantly larger decay rates for Fig.~\ref{fig:prob_fid_plot}(c), and observe obvious concave curves in the fidelity. For smaller $\gamma$, the reduced distinguishability among $k$ indices leads to a trade-off between fidelity and success probability—a hallmark of probabilistic protocols~\cite{BartkiewiczPRA13, KumarPRA23, MendozaFierroQIP26, PodoshvedovSR23, ZhaoNP20, SliwaPRA03, TsujimotoNC25, SangouardPRL11}. Conversely, for larger $\gamma$, dissipation dominates and diminishes both quantities. At the optimal point, the peak fidelity shifts slightly away from the ideal value of $k=15$ due to amplitude damping. Nevertheless, as established earlier, any sufficiently large $k$ constitutes a successful protocol outcome. Hence, this minor shift carries no practical consequence for the cat-state generation.

\section{Discussion and conclusion}
We note that the protocol naturally extends to the preparation of $d$-mode multinomial cat states for $d\geq 2$. The key observation is that, analogous to Eq.~\eqref{eq_conversion}, a product of $d$ coherent states admits the decomposition (see appendix~\ref{sec_app_detailed_calculation})
\begin{align}
    \nonumber |\alpha_1\rangle\otimes\dots\otimes|\alpha_d\rangle=\sum_{k=1}^\infty \sqrt{P(k)}|k,\xi_1,\dots,\xi_{d-1}\rangle, \label{eq_conversion_d}
\end{align}
where $|k,\xi_1,\dots,\xi_{d-1}\rangle$ denotes a $d$-mode multinomial coherent state and $P(k)$ is a Poisson distribution with mean $\lambda=\sum_{i=1}^{d}|\alpha_i|^2$. The remainder of the protocol then proceeds exactly as in the two-mode case discussed above: cross-Kerr interactions map the total excitation number $k$ of the system onto the phase of an auxiliary coherent state, and a subsequent heterodyne measurement probabilistically selects a particular value of $k$, thereby heralding the corresponding $d$-mode multinomial cat state.

More generally, our protocol is closely analogous to probabilistic Fock state preparation schemes~\cite{ZhangPRA24, DengNP24}. As shown in Eq.~\eqref{eq_conversion} and Eq.~\eqref{eq_conversion_d}, the product of several coherent states can be expressed as a Poissonian superposition in the $d$-mode multinomial state basis $|k,\xi_i\rangle$, which plays a role analogous to the Fock state $|k\rangle$. From this perspective, selecting a particular excitation index $k$ to prepare a two-mode binomial cat state is analogous to projecting a single-mode coherent state (which also follows Poissonian statistics) onto a specific Fock state $|k\rangle$. This connection suggests that techniques developed for probabilistic Fock-state generation could be adapted to further improve our protocol.

In conclusion, we have proposed a probabilistic protocol for preparing two-mode binomial cat states using cross-Kerr interactions and Gaussian input states. We show that high-fidelity ($>90\%$) cat states can be generated for sufficiently large $k$, with the accumulated success probability (density) approaching $50\%$. Moreover, the protocol can be made robust against dissipation by appropriately optimizing the coherent-state amplitude of the auxiliary mode. Our work provides a promising route towards the efficient generation of multi-mode cat states using simple state-of-the-art experimental set-ups.

\section{Acknowledgement}
SQ is funded by the Wallenberg Initiative on Networks and Quantum Information (WINQ) at Nordita, where  Nordita is partially supported by NordForsk.

\clearpage
\appendix
\section{Derivation of the protocol extended to d-mode multinomial cat states}\label{sec_app_detailed_calculation}
In this section, we provide a detailed derivation showing that the first two modes of the initial state $|\alpha\rangle\otimes|\beta\rangle\otimes|\gamma\rangle$ can evolve into a two-mode binomial cat state under the Kerr Hamiltonian in Eq.~\eqref{eq_sk_ck} at time $t_0$. Then, we show that this cat state preparation strategy can be extended to prepare $d$-mode multinomial cat states for $d\geq 2$.

To obtain Eq.~\eqref{eq_conversion}, we first assume real coherent amplitudes $\alpha,\beta$ without loss of generality, then \begin{align}
    \nonumber &|\alpha\rangle\otimes|\beta\rangle,\\
    \nonumber =\ &e^{\frac{\alpha^2+\beta^2}{2}}\sum_{n_1=0}^\infty\sum_{n_2=0}^\infty \frac{\alpha^{n_1}\beta^{n_2}}{\sqrt{n_1!n_2!}}|n_1\rangle\otimes|n_2\rangle,\\
    \nonumber =\ &e^{\frac{\alpha^2+\beta^2}{2}}\sum_{k=0}^\infty \sum_{n_1=0}^\infty \frac{\alpha^{n_1}\beta^{k-n_1}}{\sqrt{n_1!(k-n_1)!}}|n_1\rangle\otimes|k-n_1\rangle,\\
    \nonumber =\ &\sum_{k=0}^\infty e^{\frac{\alpha^2+\beta^2}{2}} \sqrt{\frac{(\alpha^2+\beta^2)^k}{k!}} \sum_{n_1=0}^k \sqrt{\frac{k!}{(1+(\alpha/\beta)^2)^k}}\\
    \nonumber &\qquad \times\frac{(\alpha/\beta)^{n_1}}{\sqrt{n_1!(k-n_1)!}}|n_1\rangle\otimes|k-n_1\rangle,\\
    =\ &\sum_{k=0}^\infty \sqrt{P(k)}|k,\xi\rangle,
\end{align}
where $P(k)=e^\lambda \lambda^k/k!$ is the Poisson distribution with mean $\lambda=\alpha^2+\beta^2$, and $|k,\xi\rangle$ is the two-mode binomial coherent state with amplitude $\xi=\alpha/\beta$.

To obtain Eq.~\eqref{Eq12}, we apply the time evolution operator $\hat{U}(t)=e^{-i\hat{H}t}$ on the initial state $|\alpha\rangle\otimes|\beta\rangle\otimes|\gamma\rangle$, where $\hat{H}$ is defined in Eq.~\eqref{Eq02} with $g_{ac}=g_{bc}$: \begin{align}
    \nonumber &e^{-i\hat{H}t}|\alpha\rangle\otimes|\beta\rangle\otimes|\gamma\rangle,\\
    \nonumber =\ &e^{-i\hat{H}t}\sum_{k=0}^\infty \sqrt{P(k)}\ |k,\xi\rangle\otimes |\gamma\rangle,\\
    \nonumber =\ &\sum_{k=0}^\infty \sqrt{P(k)}\sum_{n_1=0}^k c_k e^{-it(g_an_1^2+g_b(k-n_1)^2+g_{ab}n_1(k-n_1))}\left(\frac{\alpha}{\beta}\right)^{n_1},\\
    \nonumber &\qquad \times |n_1\rangle\otimes |k-n_1\rangle \otimes e^{-it(g_c(\hat{c}^\dag\hat{c})^2+g_{ac}k\hat{c}^\dag\hat{c})}|\gamma\rangle,\\
    \nonumber =\ &\sum_{k=0}^\infty \sqrt{P(k)} \sum_{n_1=0}^k c_k e^{-it(g_a+g_b-g_{ab})n_1^2}\left(e^{-i(g_{ab}-2g_b)kt}\frac{\alpha}{\beta}\right)^{n_1},\\
    \nonumber &\qquad \times |n_1\rangle\otimes |k-n_1\rangle \otimes e^{-ig_{ac}(\hat{c}^\dag\hat{c})^2t}|e^{-ig_{ac}kt}\gamma\rangle,\\
    =\ &\sum_{k=0}^\infty \sqrt{P(k)}\  e^{-i\mathcal{K}_2(\hat{a}^\dag\hat{a})^2t}|k,\xi_k\rangle\otimes e^{-ig_c(\hat{c}^\dag\hat{c})^2}|e^{-ig_{ac}kt}\gamma\rangle,
\end{align}
where we have use the short-hand notation $c_k=e^{-ig_{b}k^2t}\sqrt{k!/(n_1!(k-n_1)!(1+(\alpha/\beta)^2)^k)}$, $\mathcal{K}_2=g_a+g_b-g_{ab}$, and $\xi_k=(e^{-i(g_{ab}-2g_b)kt}\alpha/\beta)^{n_1}$.

At time $t_0=\pi/(2|\mathcal{K}_2|)$, the system mode (modes $a$ and $b$) becomes the cat state (see Eq.~\eqref{Eq03}) \begin{align}
    \nonumber &e^{-i\mathcal{K}_2(\hat{a}^\dag\hat{a})^2t_0}|k,\xi_k\rangle,\\
    \nonumber =\ &\frac{1}{\sqrt{\mathcal{N}}}\sum_{n_1=0}^k e^{-i\pi(\hat{a}^\dag\hat{a})^2/2}\frac{\xi_k^{n_1}}{\sqrt{n_1!(k-n_1)!}}|n_1\rangle\otimes|k-n_1\rangle,\\
    \nonumber =\ &\frac{1}{\sqrt{\mathcal{N}}}\left(\sum_{n_1=\mathrm{even}}^k-\ i\sum_{n_1=\mathrm{odd}}^k\right)\frac{\xi_k^{n_1}}{\sqrt{n_1!(k-n_1)!}}|n_1\rangle\otimes|k-n_1\rangle,\\
    \nonumber =\ &\frac{1}{\sqrt{2}}\left(\frac{1-i}{2}|k,\xi_k\rangle+\frac{1+i}{2}|k,-\xi_k\rangle\right),\\
    =\ &|\psi(k,\xi_k)\rangle
\end{align}
where $\mathcal{N}=(1+(\alpha/\beta)^2)^k/k!$ is the normalization factor, and this is exactly analogous to the single-mode case~\cite{YurkePRL86} \begin{align}
    \nonumber &e^{-i\pi(\hat{a}^\dag\hat{a})^2/2}|\alpha\rangle,\\
    \nonumber =\ &e^{-\frac{|\alpha|^2}{2}}\left(\sum_{n_1=\mathrm{even}}^k-\ i\sum_{n_1=\mathrm{odd}}^k\right)\frac{\alpha^{n_1}}{\sqrt{n_1!}}|n_1\rangle,\\
    =\ &\frac{1}{\sqrt{2}}\left(\frac{1-i}{2}|\alpha\rangle+\frac{1+i}{2}|-\alpha\rangle\right),
\end{align}
where sufficiently large $\alpha$ -- such that $\langle\alpha|-\alpha\rangle\approx 0$ -- ensures macroscopic distinguishability of the two cat-state components.

Furthermore, we may also generalize Eq.~\eqref{eq_conversion} to $d\geq2$ modes: \begin{align}
    \nonumber |\alpha_i\rangle^{\otimes_{i=1}^{d}}&=e^{-\frac{1}{2}\sum_i^d |\alpha_i|^2}\sum_{\{n_i\}}^\infty \prod_i \frac{\alpha_i^{n_i}}{\sqrt{n_i!}}|n_i\rangle^{\otimes_i^d},\\
    \nonumber &=\sum_{k=0}^\infty e^{-\frac{1}{2}\sum_i^d |\alpha_i|^2}\sqrt{\frac{(\sum_i^d\alpha_i^2)^k}{k!}}\\
    \nonumber &\qquad \times\sum_{\{n_{i<d}\}}^{\sum_{i=1}^{d-1}n_i\leq k}\sqrt{\frac{k!}{(1+\sum_{i=1}^{d-1}(\alpha_i/\alpha_d)^2)^k}}\\
    \nonumber &\qquad \times \frac{1}{\sqrt{(k-\Sigma_{i=1}^{d-1}n_i)!}}\prod_{i=1}^{d-1}\frac{(\alpha_i/\alpha_d)^{n_i}}{\sqrt{n_i!}}\\
    \nonumber &\qquad \times |n_i\rangle^{\otimes_{i=1}^{d-1}}\otimes|k-\Sigma_{i=1}^{d-1}n_i\rangle,\\
    &=\sum_{k=1}^\infty \sqrt{P(k)}|k,\xi_1,\dots,\xi_{d-1}\rangle,
\end{align}
where $P(k)$ is the Poisson distribution with mean $\lambda=\sum_{i=1}^d|\alpha_i|^2$, and \begin{align}
    \nonumber |k,\xi_1,\dots,\xi_{d-1}\rangle&=\frac{1}{\mathcal{N}}\sum_{\{n_{i<d}\}}\sqrt{\binom{k}{\{n_i\}}}\prod_{i=1}^{d-1}\xi_i^{n_i}\\
    &\qquad \times|n_i\rangle^{\otimes_{i=1}^{d-1}}\otimes|k-\Sigma_{i=1}^{d-1}n_i\rangle
\end{align}
where $\mathcal{N}=(1+\sum_{i=1}^{d-1}(\alpha_i/\alpha_d)^2)^k$, and $\xi_i=\alpha_i/\alpha_d$ is the $d$-mode multinomial coherent state.

\section{Self- and cross-Kerr interactions in superconducting devices}\label{sec_app_self_cross}
In this section, we demonstrate that self- and cross-Kerr interactions between three superconducting resonators can be induced by coupling them to a nonlinear element, such as the Superconducting Nonlinear Asymmetric Inductive Element (SNAIL) as discussed in Fig.~1.(b) of~\cite{ChapmanPQ23a}. The potential of the set-up can be expanded as \begin{align}
    \hat{H}&=E_J\left(\frac{\tilde{c}_3}{3!}\hat{\phi}^3+\frac{\tilde{c}_4}{4!}\hat{\phi}^4+\frac{\tilde{c}_5}{5!}\hat{\phi}^5+\dots\right), \label{eq_nonlinear_device_potential}
\end{align}
where $E_J$ is the Josephson energy of the larger Josephson junction in the SNAIL element, $\hat{\phi}=\sum_{\eta\in\{a,b,c\}} \phi_\eta(e^{-i\omega_\eta t}\hat{\eta}+e^{i\omega_\eta t}\hat{\eta}^\dag)$ is the total flux operator of the system, with $\phi_\eta$ being the zero-point fluctuation in mode $\eta\in\{a,b,c\}$, and $\tilde{c}_3,\ \tilde{c}_5\sim\sin(\phi_\mathrm{flux})$ and $\tilde{c}_4\sim\{\sin(\phi_\mathrm{flux}),\cos(\phi_\mathrm{flux})\}$ are expansion coefficients with $\phi_\mathrm{flux}$ being the phase difference induced by an external flux through the inductor.

First, by tuning the external flux $\sin(\phi_\mathrm{flux})=0$, we can eliminate the unwanted third $\tilde{c}_3$ and fifth $\tilde{c}_5$ terms in Eq.~\eqref{eq_nonlinear_device_potential}, so that the leading order of the Hamiltonian corresponds to only the fourth order \begin{align}
    \hat{H}&= E_J\frac{\tilde{c}_4}{4!}\left[\sum_\eta \phi_\eta(e^{-i\omega_\eta t}\hat{\eta}+e^{i\omega_\eta t}\hat{\eta}^\dag)\right]^4\raisebox{-0.3cm}{.}
\end{align}
Then, by choosing different resonance frequencies $\omega_{\eta}$ to avoid any resonance conditions, i.e., $\sum_{j=1}^4\omega_j\neq0$ for $\omega_j\in\{\pm\omega_{\eta}\}$, the surviving terms after the application of rotating wave approximation (RWA) become \begin{align}
    \hat{H}&= E_J\frac{\tilde{c}_4}{4}\left(\sum_\eta \phi_\eta^4 (\hat{\eta}^\dag\hat{\eta})^2+2\sum_{\varsigma\neq\eta}\phi_\eta^2\phi_\varsigma^2\hat{\eta}^\dag\hat{\eta}\hat{\varsigma}^\dag\hat{\varsigma}\right)+\dots, \label{Eq20}
\end{align}
which consists of self-Kerr $(\hat{\eta}^\dag\hat{\eta})^2$ and cross-Kerr $\hat{\eta}^\dag\hat{\eta}\hat{\varsigma}^\dag\hat{\varsigma}$ interactions for $\eta,\ \varsigma\in\{a,b,c\}$. The additional $+\dots$ term in Eq.~\eqref{Eq20} contains contributions from the free Hamiltonian and constant scalars as a result of operator reorderings, and is ignored from this discussion. A key feature of Eq.~\eqref{Eq20} is that self-Kerr and cross-Kerr interactions always simultaneously emerge in superconducting systems, as they are both energy conserving (or independent of the choice of resonant frequencies $\omega_\eta$).

To realize the required parameter settings for ensuring an optimal performance of the protocol, i.e., $g_{ac}=g_{bc}$ and $g_c=2m\pi/t_0$, we tune the resonator fluctuations to $\phi_a=\phi_b$ and $\phi_c^4 =8m\phi_a^4$ for some interger $m$. With this, other dependent parameters become $g_a=g_b=E_J\frac{\tilde{c}_4}{4}\phi_a^4$, $g_{ac}=g_{bc}=\sqrt{8m}E_J\tilde{c}_4\phi_a^4$, $g_{ab}=E_J\tilde{c}_4\phi_a^4$, and finally $g_c=E_J\frac{\tilde{c}_4}{4}\phi_c^4=2mE_J\tilde{c}_4\phi_a^4=\frac{2m\pi}{t_0}$.

\section{Kerr evolution in a noisy environment}\label{sec_app_open_system_evolution}
In this section, we study the open-system dynamics of a three-mode bosonic system governed by the Hamiltonian in Eq.~\eqref{eq_sk_ck}, incorporating thermalization, single-photon loss, and dephasing. Using the Lie-algebraic decoupling method~\cite{WeiJMP63, ChaturvediJMO91, ChaturvediPRA91,qvarfort2025solving}, we show that the full dynamics can be decomposed into a product of distinct sub-dynamics, providing a transparent description of the individual contributions from the different physical processes.

\subsection{Solution to the Lindblad equation}\label{subsec_solution_to_lindblad}
We consider the Lindblad master equation
\begin{align}
    \dot{\hat{\rho}} = \hat{\mathcal{L}}\hat{\rho},
\end{align}
where the Liouvillian $\hat{\mathcal{L}}$ includes both the coherent evolution generated by the Hamiltonian in Eq.~\eqref{eq_sk_ck}, and the dissipative processes. Specifically, each bosonic mode $\eta$ is coupled to a thermal environment through single-photon loss and gain, described by the Lindblad operators
\begin{align}
    \hat{L}_{1,-}^{(\eta)} &= \sqrt{(1+\bar{n}^{(\eta)})\kappa_{1}^{(\eta)}}\,\hat{\eta},\\
    \hat{L}_{1,+}^{(\eta)} &= \sqrt{\bar{n}^{(\eta)}\kappa_{1}^{(\eta)}}\,\hat{\eta}^\dag,
\end{align}
respectively, together with a pure-dephasing channel described by
\begin{align}
    \hat{L}_{2}^{(\eta)} = \sqrt{\kappa_{2}^{(\eta)}}\,\hat{\eta}^\dag\hat{\eta}.
\end{align}
Here, $\kappa_{1}^{(\eta)}$ denotes the single-photon decay rate of mode $\eta$, $\bar{n}^{(\eta)}$ is the mean thermal occupation of its environment, and $\kappa_{2}^{(\eta)}$ is the corresponding pure-dephasing rate. The first two Lindblad operators account for thermal-photon loss and excitation, while the third describes dephasing without energy exchange.

In a vectorized formalism~\cite{QvarfortPRA21}, the above Lindblad equation can be rewritten as \begin{align}
    \nonumber |\dot{\hat{\rho}}\rrangle&=\bigg[\sum_{\eta}-\left(ig_\eta\hat{K}_{0}^{(\eta)}+i\sum_{\varsigma\neq\eta}\frac{g_{\eta\varsigma}}{2}\hat{K}_{0,\varsigma}+\frac{\kappa_{1}^{(\eta)}}{2}(2\bar{n}^{(\eta)}+1)\right)\hat{K}_{3}^{(\eta)}\\
    \nonumber &\hspace{1cm}-\frac{\kappa_{2}^{(\eta)}}{2}\hat{K}_{0}^{(\eta)2}+\kappa_{1}^{(\eta)}(\bar{n}^{(\eta)}+1)\hat{K}_{-}^{(\eta)}\\
    &\hspace{1cm}+\kappa_{1}^{(\eta)}\bar{n}^{(\eta)}\hat{K}_{+}^{(\eta)}-\kappa_{1}^{(\eta)}\bar{n}^{(\eta)}\bigg]|\hat{\rho}\rrangle, \label{Eq04}
\end{align}
where \begin{gather}
    \nonumber \hat{K}_{-}^{(\eta)}=\hat{\eta}\otimes\hat{\eta}, \hspace{0.8cm}\hat{K}_{+}^{(\eta)}=\hat{\eta}^\dag\otimes\hat{\eta}^\dag,\\
    \nonumber \hat{K}_{3}^{(\eta)}=\hat{\eta}^\dag\hat{\eta}\otimes\mathds{1}+\mathds{1}\otimes\hat{\eta}^\dag\hat{\eta}, \hspace{0.8cm}\hat{K}_{0}^{(\eta)}=\hat{\eta}^\dag\hat{\eta}\otimes\mathds{1}-\mathds{1}\otimes\hat{\eta}^\dag\hat{\eta},
\end{gather}
and $\hat{K}_{0}^{(\eta)}$ is the Casimir element that commutes with all other generators, which encodes the symmetry of the dynamics~\cite{AlbertPRA14a}.

The solution to the master equation Eq.~\eqref{Eq04} is~\cite{ChaturvediJMO91, ChaturvediPRA91} \begin{align}
    |\hat{\rho}(t)\rrangle&=\hat{\mathcal{S}}_C(t)\hat{\mathcal{S}}(t)|\hat{\rho}(0)\rrangle, \label{Eq14}
\end{align}
where the dynamics $\hat{\mathcal{S}}_{C}$ generated by the Casimir elements can be conveniently factored out as \begin{align}
    \hat{\mathcal{S}}_C(t)&=e^{-\frac{\kappa_{2}^{(a)}t}{2}\hat{K}_{0}^{(a)2}}e^{-\frac{\kappa_{2}^{(b)}t}{2}\hat{K}_{0}^{(b)2}}e^{-\frac{\kappa_{2}^{(c)}t}{2}\hat{K}_{0}^{(c)2}}, \label{Eq05}
\end{align}
and \begin{align}
    \hat{\mathcal{S}}(t)&=e^{\hat{F}_c}\prod_\eta e^{\hat{F}_{+}^{(\eta)}\hat{K}_{+}^{(\eta)}}\prod_\eta e^{\hat{F}_{3}^{(\eta)}\hat{K}_{3}^{(\eta)}}\prod_\eta e^{\hat{F}_{-}^{(\eta)}\hat{K}_{-}^{(\eta)}}. \label{Eq06}
\end{align}
Here, $F_\bullet$ are time-dependent coefficients defined as \begin{align}
    \nonumber \hat{F}_{+}^{(\eta)}&=\frac{2\kappa_{1}^{(\eta)}\bar{n}^{(\eta)}\tanh(\frac{\hat{B}_\eta}{2}t)}{\hat{B}_\eta+(\hat{A}_\eta+\kappa_{1}^{(\eta)}(2\bar{n}^{(\eta)}+1))\tanh(\frac{\hat{B}_\eta}{2}t)}\raisebox{-0.3cm}{,}\\
    \nonumber \hat{F}_{3}^{(\eta)}&=-\ln(\cosh(\frac{\hat{B}_\eta}{2}t)+\frac{\hat{A}_\eta+\kappa_{1}^{(\eta)}(2\bar{n}^{(\eta)}+1)}{\hat{B}_\eta}\sinh(\frac{\hat{B}_\eta}{2}t))\raisebox{-0.3cm}{,}\\
    \nonumber \hat{F}_{-}^{(\eta)}&=\frac{2\kappa_{1}^{(\eta)}(\bar{n}^{(\eta)}+1)\tanh(\frac{\hat{B}_\eta}{2}t)}{\hat{B}_\eta+(\hat{A}_\eta+\kappa_{1}^{(\eta)}(2\bar{n}^{(\eta)}+1))\tanh(\frac{\hat{B}_\eta}{2}t)}\raisebox{-0.3cm}{,}\\
    \hat{F}_c&=\sum_\eta \hat{F}_{3}^{(\eta)}+\frac{1}{2}\hat{A}_\eta t+\frac{\kappa_{1}^{(\eta)}}{2}t, \label{Eq16}
\end{align}
where \begin{align}
    \nonumber \hat{A}_\eta&=2ig_\eta\hat{K}_{0}^{(\eta)}+i\sum_{\varsigma\neq\eta}g_{\eta\varsigma}\hat{K}_{0,\varsigma},\\
    \hat{B}_\eta&=\sqrt{(\hat{A}_\eta+\kappa_{1}^{(\eta)})^2+4\hat{A}_\eta\kappa_{1}^{(\eta)}\bar{n}^{(\eta)}}.
\end{align}

\subsection{Success probability and fidelity}\label{subsec_probability_and_fidelity}
Having obtained the dissipative evolution for the system, we now compute the success probability Eq.~\eqref{eq_probability_special_case} and fidelity Eq.~\eqref{eq_fidelity_special_case} for the special case ($\kappa_2^{(\eta)}=\bar{n}=0$) considered in the main context. We first note that in this case, the system evolves as \begin{align}
    \hat{\mathcal{S}}(t)&=\prod_{\eta}e^{-\frac{1}{2}(\hat{A}_\eta+\kappa_1^{(\eta)})\hat{K}_3^{(\eta)}t}\prod_\eta e^{\hat{F}_-^{(\eta)}\hat{K}_-^{(\eta)}},
\end{align}
where $\hat{F}_{-}^{(\eta)}=\frac{\kappa_\eta}{\hat{A}_\eta+\kappa_{\eta}}(1-e^{-(\hat{A}_{\eta}+\kappa_{\eta})t})$. With this, an initially pure state $|\psi_0\rangle=|\alpha\rangle\otimes|\beta\rangle\otimes|\gamma\rangle$ evolves as \begin{align}
    \nonumber \hat{\rho}(t_0)&=\hat{\mathcal{S}}(t_0)|\psi_0\rrangle,\\
    \nonumber &=\sum_{k_1,k_2}\prod_\eta e^{|\nu_\eta|^2\hat{F}_-} e^{-|\nu_\eta|^2+|\nu_\eta'|^2}e^{-\hat{A}_\eta\hat{K}_3^{(\eta)}}|\alpha',\beta',\gamma'\rrangle,\\
    \nonumber &=\sum_{k_1,k_2}  f(\hat{F}_-)\sqrt{P'(k_1)P'(k_2)}|\psi(k_1,\xi_{k_1}'), \psi(k_2,\xi_{k_2}')\rangle\\
    &\qquad \otimes|\gamma_{k_1}',\gamma_{k_2}'\rangle,
\end{align}
where we have used the short-handed denotation $|\psi,\phi\rangle\equiv|\psi\rangle\otimes|\phi\rangle$, $P(k)=e^{-\lambda}\lambda^k/k!$ is the Poisson distribution, and defined the decoherence quantity \begin{align}
    f(\hat{F}_-)=\prod_{\eta=a,b,c}e^{(\hat{F}_{-}^{(\eta)}-1+e^{-\kappa_{\eta} t})|\nu_\eta|^2},
\end{align}
using the damped amplitudes $\nu_\eta'=\nu_\eta e^{-\kappa_\eta t_0/2}$ for $\nu_{\eta\in{a,b,c}}=\alpha,\beta,\gamma$. $P'(k)$ and $|\psi(k,\xi_k')\rangle$ are the Poisson distribution and two-mode binomial cat state generated using damped amplitudes.

The success probability evaluates to
\begin{align}
    \nonumber Q_\rho(k)&=\mathrm{Tr}[\hat{M}_k\hat{\rho}(t_0)\hat{M}_k^\dag],\\
    \nonumber &=\mathrm{Tr}[\hat{M}_k\hat{\rho}(t_0)],\\
    \nonumber &=\llangle \mathds{1}\otimes\mathds{1}\otimes\hat{M}_k|\hat{\rho}(t_0)\rrangle ,\\
    \nonumber &=\sum_{\{\bullet\}} \langle n_1',n_2',n_3'|\otimes\langle n_1',n_2',m_3'|f(\hat{F}_-)\sqrt{P'(k_1)P'(k_2)}\\
    \nonumber \times &\sqrt{B'(k_1,n_1)B'(k_2,n_2)}p^*(n_3')p(m_3')p'(n_3)p'^*(m_3)\\
    \nonumber \times &e^{-ig_{ac}(k_1-k)(n_3-m_3)t_0}|n_1,k_1-n_1, n_3\rangle\otimes|n_2,k_2-n_2, m_3\rangle,\\
    \nonumber &=\sum_{\substack{n_1,n_3\\ m_3,k_1}}f_\mathrm{prob}(n_3,m_3)P'(k_1)B'(k_1,n_1)p^*(n_3)\\
    \times p&(m_3)p'(n_3)p'^*(m_3)e^{-ig_{ac}(k_1-k)(n_3-m_3)t_0}, \label{eq_app_success_probability_special_case}
\end{align}
where the summation over $\{\bullet\}$ conveniently indicates summing over all subsequent indices, $B'(k_1,n_1)=\binom{k_1}{n_1}|\frac{\alpha'}{\beta'}|^{2n_1}(1+|\frac{\alpha'}{\beta'}|^2)^{-k_1}$ is the binomial distribution generated by the damped amplitudes, $p(n)=e^{-\frac{|\gamma|^2}{2}}\frac{\gamma^n}{\sqrt{n!}}$ and $p'(n)=e^{-\frac{|\gamma'|^2}{2}}\frac{\gamma'^n}{\sqrt{n!}}$ are Poisson distribution amplitudes.

Similarly, the fidelity is
\begin{align}
    \nonumber \mathcal{F}(k)&=\frac{\langle\psi(k,\xi_k)|\mathrm{Tr}_c[\hat{M}_k\hat{\rho}(t_0)\hat{M}_k^\dag]|\psi(k,\xi_k)\rangle}{Q_\rho(k)},\\
    \nonumber &=\frac{\langle\psi(k,\xi_k)|\mathrm{Tr}_c[\hat{M}_k\hat{\rho}(t_0)]|\psi(k,\xi_k)\rangle}{Q_\rho(k)},\\
    \nonumber &=\frac{1}{Q_\rho(k)}\llangle \hat{\rho}_t\otimes\hat{M}_k|\hat{\rho}(t_0)\rrangle,\\
    \nonumber &=\frac{1}{Q_\rho(k)}\llangle \hat{\rho}_k\otimes\hat{M}_k|\hat{\mathcal{U}}^\dag|\hat{\rho}(t_0)\rrangle,\\
    \nonumber &= \frac{1}{Q_\rho(k)} \sum_{k_1,k_2} \llangle \hat{\rho}_k\otimes\hat{M}_k| f(\hat{F}_-)\sqrt{P'(k_1)P'(k_2)}\\
    \nonumber \times &|(k_1,\xi_{k_1}'), (k_2,\xi_{k_2}')\rangle\otimes |\gamma_{k_1}',\gamma_{k_2}'\rangle \\
    \nonumber &=\frac{P'(k)}{Q_\rho(k)}\sum_{\substack{n_1,n_3\\ m_1,m_3}}f_\mathrm{fid}(n_1,m_1,n_3,m_3)\\
    \nonumber&\qquad \times b^*(k,n_1)b(k,m_1)b'(k,n_1)b'^*(k,m_1)\\
    &\qquad \times p^*(n_3)p(m_3)p'(n_3)p'^*(m_3),\label{eq_app_fidelity_special_case}
\end{align}
where $\hat{\mathcal{U}}=\hat{U}\otimes\hat{U}^\dag$ is the self-Kerr evolution operator that forms the target cat state, $\hat{\rho}_t=|\psi(k,\xi_k)\rangle\langle\psi(k,\xi_k)|$ and $\hat{\rho}_k=|k,\xi_k\rangle\langle k,\xi_k|$ denote the projectors onto the two-mode binomial cat state and coherent state respectively, $b(k,n)=\binom{k}{n}^{\frac{1}{2}}\xi^n(1+|\xi|^2)^{-\frac{k}{2}}$ and $b'(k,n)=\binom{k}{n}^{\frac{1}{2}}(\frac{\alpha'}{\beta'})^n(1+|\frac{\alpha'}{\beta'}|^2)^{-\frac{k}{2}}$ are complex binomial distribution amplitudes. We note that in the fourth line, we replaced the cat state in the bra with the Kerr-evolved coherent state $|\psi(k,\xi_k)\rangle=\hat{U}|k,\xi_k\rangle$. Since the corresponding superoperator $\hat{\mathcal{U}}$ commutes with $f(\hat{F}_-)$, its inverse operation can be equivalently applied to the cat state in the ket, reverting it back to the original coherent state. We note that Eq.~\eqref{eq_app_fidelity_special_case} involves multiple nested summations, and its numerical evaluation can become computationally demanding for larger values of $\gamma$, as the corresponding summation ranges increase.

\subsection{Conditions for maintaining quantum purity}\label{subsec_conditions_for_maintaining_purity}
After decomposing the full dynamics into distinct sub-dynamics as shown in Eq.~\eqref{Eq14}, we can identify the individual contributions responsible for the loss of state purity. In particular, we show that a superoperator with a nonlinear exponential dependence on the Casimir element $\hat{K}_0$ introduced above, of the form $e^{f(\hat{K}_0)}$ with $f(\hat{K}_0)$ a nonlinear function in $\hat{K}_0$, necessarily reduces the purity of the quantum state. For instance, the dephasing-induced dynamics in Eq.~\eqref{Eq05} explicitly contains a term of the form $e^{\hat{K}_0^2}$, and hence directly contributes to the reduction of the system purity. For the remaining dynamical terms, which may contain more complicated dependence on $\hat{K}_0$, we can expand the corresponding functions in appropriate limits to separate their linear and nonlinear contributions.

More specifically, in the Fock basis, we consider two classes of functional dependence: $f(\hat{K}_0)\equiv f(m-n)$ and $f(\hat{K}_0)^{\hat{K}_3}\equiv f(m-n)^{m+n}$, where $\hat{K}_3\hat{\rho}=\hat{a}^\dag\hat{a}\hat{\rho}+\hat{\rho}\hat{a}^\dag\hat{a}$. Since we assume a pure initial state throughout this work, the corresponding functionals must preserve the idempotency condition $\hat{\rho}^2=\hat{\rho}$. The first functional $f(\hat{K}_0)$ induces the equation \begin{align}
    \sum_{j}|c_{j,j}|^2f(m-j)f(j-n)&=f(m-n), \label{Eq18}
\end{align}
for which we can prove that its unique solution is $f(\hat{K}_0)=e^{z\hat{K}_0}$. We first claim that $f(m-j)f(j-m)$ is independent of $j$ to obtain the equation \begin{align}
    \nonumber f(m-j)f(j-n)&=f(m-n),\\
    e^{h(m-j)+h(j-n)}&=e^{h(m-n)}.
\end{align}
Since $h(n)$ is an odd function, the only solution is $h(n)=zn$ for $z\in\mathds{C}$, and thus $f(\hat{K}_0)=e^{z\hat{K}_0}$. We prove our claim by contradiction. Suppose $f(m-j)f(j-m)$ depend on $j$, and without loss of generality, we can assume that they are linearly independent. By identifying Eq.~\eqref{Eq18} as a convex combination, we observe that a shift in $m\rightarrow m+1$ and $n\rightarrow n+1$ induces a shift in $j\rightarrow j-1$. This suggests that there are two sets of parameters $|c_{j,j}|^2$ and $|d_{j,j}|^2=|c_{j+1,j+1}|^2$ that induce the same point on the affine plane, which contradicts the uniqueness of convex combination.

Similarly, the second functional $f(\hat{K}_0)^{\hat{K}_3}$ induces the equation \begin{align}
    \sum_{j}|c_{j,j}|^2f(m-j)^{m+j}f(j-n)^{j+n}&=f(m-n)^{m+n}.
\end{align}
We similarly claim that $f(m-j)^{m+j}f(j-n)^{j+n}$ is independent of $j$ to obtain the equation \begin{align}
    e^{h(m-j)(m+j)+h(j-n)(n+j)}&=e^{h(m-n)(m+n)},
\end{align}
which restricts the solution to $h(n)=zn$, and hence $f(\hat{K}_0)^{\hat{K}_3}=e^{z\hat{K}_0\hat{K}_3}$ becomes the only solution. We prove our claim by contradiction. Suppose $f(m-j)^{m+j}f(j-n)^{j+n}$ depends on $j$. Since $m$, $n$ and $j$ are no longer dependent parameters, we can obtain a matrix equation by identifying $m$ and $n$ as the entry indices of the matrix $M_{m,n}$: \begin{align}
    M&=M\Sigma M, \label{Eq21}
\end{align}
where $\Sigma_{m,m}=|c_{m,m}|^2$ is a diagonal matrix, and $M_{m,n}=f(m-n)^{m+n}$. We observe that both $\Sigma$ and $M$ are invertible. Hence, the unique solution to Eq.~\eqref{Eq21} is $M=\Sigma^{-1}$, which contradicts our assumption that $M$ has dependence on $\hat{K}_0$ and $\hat{K}_3$.

Moreover, since the purity function $\Tr[(f(\hat{K}_0)\hat{\rho})^2]$ is continuous over the convex set of density matrices, the above results derived under the assumption of an initially pure state naturally extend to arbitrary mixed states.

\subsection{Expansion of \texorpdfstring{$\hat{F}_{3}^{(\eta)}$}{}}\label{subsec_expansion_f3}
Having identified the functional forms involving $\hat{K}_0$ that preserve the system purity, we can now use this insight to analyze the more complicated sub-dynamics in Eq.~\eqref{Eq06}. By expanding these terms in appropriate limits, we can isolate their linear and nonlinear dependence on $\hat{K}_0$ in the exponential, and thereby identify their respective effects on the system state. We begin by expanding $\hat{F}_{3}^{(\eta)}$, which allows us to separate its contributions to the coherent amplitude damping and purity loss.

We first note that since $e^{\hat{F}_{3}^{(\eta)}\hat{K}_3}|\alpha\rrangle=|e^{\hat{F}_3^{(\eta)}}\alpha\rangle$, it suffices to analyze the exponential factor \begin{align}
    e^{\hat{F}_{3}^{(\eta)}} &=\left(\cosh(\frac{\hat{B}_\eta}{2}t)+\frac{\hat{A}_\eta+\kappa_{1}^{(\eta)}(2\bar{n}^{(\eta)}+1)}{\hat{B}_\eta}\sinh(\frac{\hat{B}_\eta}{2}t)\right)^{-1}.
\end{align}
Then, we can rewrite the fraction prefactor as \begin{align}
    \nonumber \frac{\hat{A}_\eta+\kappa_{1}^{(\eta)}(2\bar{n}^{(\eta)}+1)}{\hat{B}_\eta}&=\left(\frac{\hat{B}_\eta^2+4\kappa_{1}^{(\eta)2}\bar{n}^{(\eta)}(\bar{n}^{(\eta)}+1)}{\hat{B}_\eta^2}\right)^\frac{1}{2}\raisebox{-0.3cm}{,}\\
    \nonumber &=\left(1+\frac{4\kappa_{1}^{(\eta)2}\bar{n}^{(\eta)}(\bar{n}^{(\eta)}+1)}{\hat{B}_\eta^2}\right)^{\frac{1}{2}}\raisebox{-0.3cm}{,}\\
    &=1+\frac{2\kappa_{1}^{(\eta)2}\bar{n}^{(\eta)}(\bar{n}^{(\eta)}+1)}{\hat{B}_\eta^2}+...,
\end{align}
in the limit of small $\|2\kappa_{1}^{(\eta)2}\bar{n}^{(\eta)}(\bar{n}^{(\eta)}+1)/\hat{B}_\eta^2\|\ll 1$, i.e., sufficiently weak noise $\kappa_1^{(\eta)}\bar{n}^{(\eta)}$ or large self- and cross-Kerr coefficiently $g_\bullet$. Substituting this expansion back, we get \begin{align}
    \nonumber e^{\hat{F}_{3}^{(\eta)}}&=\left(e^{\frac{\hat{B}_\eta}{2}t}+\frac{2\kappa_{1}^{(\eta)2}\bar{n}^{(\eta)}(\bar{n}^{(\eta)}+1)}{\hat{B}_\eta^2}\sinh(\frac{\hat{B}_\eta}{2}t)\right)^{-1}\raisebox{-0.3cm}{,}\\
    &=e^{-\frac{\hat{B}_\eta}{2}t}\left(1-\frac{2\kappa_{1}^{(\eta)2}\bar{n}^{(\eta)}(\bar{n}^{(\eta)}+1)}{2\hat{B}_\eta^2}(1-e^{-\hat{B}_\eta t})+\dots\right)\raisebox{-0.3cm}{.}
\end{align}
Here, we can further expand \begin{align}
    \nonumber \hat{B}_\eta&=\hat{A}_\eta\left(1+\frac{\kappa_{1}^{(\eta)2}+2\hat{A}_\eta\kappa_{1}^{(\eta)}(1+2\bar{n}^{(\eta)})}{\hat{A}_\eta^2}\right)^{\frac{1}{2}}\raisebox{-0.3cm}{,}\\
    &=\hat{A}_\eta+\kappa_{1}^{(\eta)}(1+2\bar{n}^{(\eta)})+\mathcal{O}(\kappa_{1}^{(\eta)3}, \hat{A}_{\eta}^{k\neq 1}),
\end{align}
which allows us to approximate \begin{align}
    \nonumber e^{\hat{F}_{3}^{(\eta)}}&\approx e^{-\frac{\hat{A}_\eta}{2}t}e^{-\frac{\kappa_{1}^{(\eta)}(1+2\bar{n}^{(\eta)})}{2}t}\\
    &\qquad \times\left(1-\frac{2\kappa_{1}^{(\eta)2}\bar{n}^{(\eta)}(\bar{n}^{(\eta)}+1)}{2\hat{B}_\eta^2}(1-e^{-\hat{B}_\eta t})\right)\raisebox{-0.3cm}{,}
\end{align}
where the first exponential term corresponds to the phase shift due to the presence of single photon loss, and the second and third terms correspond to the amplitude drop of the coherent state.

\subsection{Expansion of \texorpdfstring{$\hat{F}_{\pm}^{(\eta)}$}{}}\label{subsec_expansion_fpm}
Similarly, we can also expand $\hat{F}_{\pm}^{(\eta)}$ defined in Eq.~\eqref{Eq16} as a power series to extract the unitary evolution induced by these terms. Since $\hat{F}_{\pm}^{(\eta)}$ differs only by a scalar factor, it suffices to showcase only one. Here, we stress that the denominator of $\hat{F}_{\pm}^{(\eta)}$ may fail to satisfy the criterion for expanding expressions of the form $\frac{1}{1+x}$ \begin{align}
    \norm{(\hat{B}_\eta)^{-1}(\hat{A}_\eta+\kappa_{1}^{(\eta)}(2\bar{n}^{(\eta)}+1))\tanh(\frac{\hat{B}_\eta}{2}t)}< 1, \label{eq_criterion}
\end{align}
so our following expansion is only valid for sufficiently small time $t=t'$ at which Eq.~\eqref{eq_criterion} is fulfilled. With this, we obtain \begin{align}
    \hat{F}_{\pm}^{(\eta)}(t)=\hat{F}_{\pm}^{(\eta)}(t')+\hat{F}_{\pm}'^{(\eta)}(t')(t-t')+...,
\end{align}
and the choice of $t'$ allows us to further expand the zeroth order term $\hat{F}_{\pm}^{(\eta)}(t')$. We note that the same procedure can also be employed to expand higher order terms such as $\hat{F}_{\pm}'^{(\eta)}$.

Up to the zeroth order $\hat{F}_{-}^{(\eta)}(t)$ can be expanded and split into two parts
\begin{align}
    \nonumber \hat{F}_{-}^{(\eta)}(t)&\approx D_{-}^{(\eta)}+\phi_{-}^{(\eta)}\hat{A}_\eta,
\end{align}
with a damping term \begin{align}
    D_{-}^{(\eta)}&=\kappa_{1}^{(\eta)}(\bar{n}^{(\eta)}+1)t'\bigg[1-\frac{1}{2}\kappa_{1}^{(\eta)}t'(2\bar{n}^{(\eta)}+1)\bigg]\raisebox{-0.3cm}{,} \label{eq_app_d_-}
\end{align}
and a unitary term \begin{align}
    \phi_{-}^{(\eta)}&=-\frac{1}{2}\kappa_{1}^{(\eta)}(\bar{n}^{(\eta)}+1)t'^2\bigg[1-\frac{2}{3}\kappa_{1}^{(\eta)}(2\bar{n}^{(\eta)}+1)t'\bigg]\raisebox{-0.3cm}{.} \label{eq_app_phi_-}
\end{align}
In particular, this unitary term indicates that the single-photon loss/gain can interplay with the self- and cross-Kerr interactions to induce an additional free evolution on the system $\hat{U}_-=\prod_\eta e^{|\nu_\eta|^2\phi_-^{(\eta)}\hat{A}_\eta}$.

Similarly, for $\hat{F}_{+}^{(\eta)}(t')$, we have \begin{align}
    \nonumber \hat{F}_{+}^{(\eta)}(t')&\approx D_{+}^{(\eta)}+ {\phi_{+}^{(\eta)}}\hat{A}_\eta,
\end{align}
where \begin{align}
    D_{+}^{(\eta)}=\frac{\bar{n}^{(\eta)}}{\bar{n}^{(\eta)}+1} D_-^{(\eta)}, \qquad \phi_{+}^{(\eta)}=\frac{\bar{n}^{(\eta)}}{\bar{n}^{(\eta)}+1}\phi_{-,\eta}.
\end{align}
Using the approximation $\hat{a}^\dag|\alpha\rangle\approx \alpha^*|\alpha\rangle$ for sufficiently large amplitude $|\alpha|\gg1$, we can similarly obtain the additional thermalization-induced free evolution $\hat{U}_+=\prod_\eta e^{|\nu_\eta|^2\phi_+^{(\eta)}\hat{A}_\eta}$.

For the special case when $\bar{n}^{(\eta)}=0$, and the limit of sufficiently small time $\norm{(\hat{A}_\eta+\kappa_{1}^{(\eta)})t'}\ll 1$, $\hat{F}_{-}^{(\eta)}$ can be expanded more accurately as \begin{align}
    \hat{F}_{-}^{(\eta)}=\kappa_{1}^{(\eta)}t'-\frac{\kappa_{1}^{(\eta)}t'^2}{2}(\hat{A}_\eta+\kappa_{1}^{(\eta)})+\dots, \label{eq_app_f_-_n_eq_0}
\end{align}
where the unitary term $-\frac{\kappa_{1}^{(\eta)}t_0^2}{2}\hat{A}_\eta$ indicates the dissipation-induced free evolution. We note that Eq.~\eqref{eq_app_f_-_n_eq_0} can alternatively be obtained by directly setting $\bar{n}=0$ in Eq.~\eqref{eq_app_d_-} and Eq.~\eqref{eq_app_phi_-}. Although the required convergence condition cannot be fulfilled in our protocol, for $t'\sim\frac{1}{g_a}$ and $\norm{\hat{A}_\eta t_0}\sim|\alpha|^2+|\beta|^2+|\gamma|^2 \gg 1$, this observation could be found useful in other protocol scenarios, such as that of in~\cite{MogilevtsevPRA09}.

\section{Further modifications on the protocol}\label{sec_app_further_modifications}
In this section, we examine several possible modifications to the protocol, and show why they do not provide a meaningful improvement in its overall performance.

First, Eq.~\eqref{eq_app_success_probability_special_case} suggests that using a corrected projective measurement $\hat{M}_k'=|\gamma_k'\rangle\langle\gamma_k'|$ could increase each summand of $Q_\rho(k)$ by the factor
\begin{align}
    G_\mathrm{prob}(n_3,m_3)=e^{(1-e^{-\kappa_{1}^{(c)}t_0})|\gamma|^2-\frac{\kappa_{1}^{(c)}}{2}(n_3+m_3)t_0}. \label{eq_gain_prob_summands}
\end{align}
For this factor to provide a positive gain $G_\mathrm{prob}(n_3,m_3)\geq 1$ over the dominant contributions to the summation, a sufficiently large $|\gamma|$ is required. However, in this regime, the fidelity can be severely harmed by the decoherence factor $f_\mathrm{fid}(\dots)$ in Eq.~\eqref{eq_app_fidelity_special_case}. Conversely, for moderate $|\gamma|$ and $\kappa_{1}^{(c)}$, where the fidelity can remain at a high level, e.g., $\sim90\%$, we find $G_\mathrm{prob}\approx1$. Thus, the corrected measurement provides essentially no appreciable improvement in the success probability in the parameter regime where the protocol maintains high fidelity.

Second, one may instead attempt to increase the fidelity by modifying the measurement amplitude, i.e., by using $\hat{M}_k(\delta_k)=|\gamma_k+\delta_k\rangle\langle\gamma_k+\delta_k|$. Such an improvement, however, comes at the cost of a reduced success probability, reflecting the typical trade-off between fidelity and success probability in probabilistic state-preparation protocols~\cite{BartkiewiczPRA13, KumarPRA23, MendozaFierroQIP26, PodoshvedovSR23, TsujimotoNC25, ZhaoNP20}. This trade-off can also be understood directly from Eq.~\eqref{eq_app_success_probability_special_case} and Eq.~\eqref{eq_app_fidelity_special_case}. Increasing the measurement amplitude from $\gamma_k$ to $\gamma_k+\delta_k$ suppresses both summations $\sum_{\substack{n_1,n_3\\ m_3,k_1}}$ and $\sum_{\substack{n_1,n_3\\ m_1,m_3}}$ in $Q_\rho(k)$ and $\mathcal{F}(k)$ respectively, with the former decreasing more rapidly. Consequently, the ratio $\mathcal{F}(k)\sim\frac{\mathrm{latter}}{\mathrm{former}}$ increases, leading to a fidelity improvement at the expense of the success probability. Physically, a larger measurement amplitude shifts the dominant support of the summands towards larger values of $n_3$ and $m_3$, causing the additional phase factor $e^{-ig_{ac}(k_1-k)(n_3-m_3)t_0}$ in the former summation to oscillate more rapidly. This produces stronger destructive interference and thereby suppresses the success probability. Thus, although this modification can improve the fidelity, the rather faster associated reduction in success probability makes it an unfavorable strategy for improving the overall protocol performance.

Third, one may consider modifying the initial coherent amplitudes $\alpha$ and $\beta$ such that the ratio $\frac{\alpha'}{\beta'}=\xi_k'$ of their damped amplitudes at time $t_0$ matches that of the target cat state $\xi_k=\xi_k'$, with the aim of improving the protocol fidelity. However, this amplitude matching does not necessarily lead to an improvement. On the one hand, increasing $|\alpha|$ and $|\beta|$ enhances the decoherence captured by the factor $f_\mathrm{fid}(\dots)$, which can outweigh any benefit from improved amplitude matching and thereby reduce the overall fidelity. On the other hand, decreasing the initial amplitudes shifts the binomial distribution $b'(k,n_1)b'^*(k,m_1)$ toward a smaller mean, thereby reducing its overlap with the target distribution $b^*(k,n_1)b(k,m_1)$, and consequently lowering the fidelity. Therefore, neither increasing nor decreasing the initial coherent amplitudes provides a systematic improvement, and the optimal strategy is to retain the original values of $\alpha$ and $\beta$.

\end{document}